\documentclass[10pt]{olplainarticle}
\usepackage[10pt]{moresize}
\usepackage{gensymb}
\usepackage{multirow}
\usepackage{amssymb,amsfonts,amsmath}
\usepackage{longtable}
\usepackage[colorinlistoftodos]{todonotes}
\usepackage[colorlinks=true, allcolors=blue]{hyperref}
\usepackage{amssymb,amsfonts}
\usepackage{cite}
\usepackage{placeins}

\DeclareMathSymbol{*}{\mathbin}{symbols}{"01} % makes dots instead of * in math mode

\usepackage[LGRgreek]{mathastext} % makes math mode non-italic
\usepackage{newfloat} % supplementary figures as a different class 
\DeclareFloatingEnvironment[name={SI Figure}]{suppfigure}

\title{\large  Single-cell bacterial electrophysiology reveals mechanisms of stress induced damage}

\author[1]{Ekaterina Krasnopeeva}
\author[2]{Chien-Jung Lo}
\author[1]{Teuta Pilizota}
\affil[1]{Centre for Synthetic and Systems Biology, Institute of Cell Biology, School of Biological Sciences, University of Edinburgh, Alexander Crum Brown Road, EH9 3FF, Edinburgh, UK}
\affil[2]{Department of Physics and Graduate Institute of Biophysics, National Central University, Jhongli, Taiwan 32001, ROC}


\begin{abstract}
\textbf{Electrochemical gradient of protons, or proton motive force (PMF), is at the basis of bacterial energetics. It powers vital cellular processes and defines the physiological state of the cell. Here we use an electric circuit analogy of an \textit{Escherichia coli} cell to mathematically describe the relationship between bacterial PMF, electric properties of the cell membrane and catabolism. We combine the analogy with the use of bacterial flagellar motor as a single-cell "voltmeter" to measure cellular PMF in varied and dynamic external environments, for example, under different stresses. We find that butanol acts as an ionophore, and functionally characterise membrane damage caused by the light of shorter wavelengths. Our approach coalesces non-invasive and fast single-cell voltmeter with a well-defined mathematical framework to enable quantitative bacterial electrophysiology.} 
\end{abstract}

\begin{document}

\flushbottom
\maketitle
\thispagestyle{empty}

\section*{Introduction}

To stay alive bacteria, like other cells, maintain adequate supplies of free energy, and under various external stresses attempt to stay viable by distributing it to processes essential for coping with the challenge, while simultaneously maintaining core cellular functions. The two main sources of free energy in living cells are adenosine triphosphate (ATP) molecule and proton motive force (PMF). The ATP molecule is the energy "currency" of living organisms used for biosynthesis and transport. The PMF is a direct consequence of the activity of the electron transport chain or substrate level phosphorylation, and serves as the energy source driving numerous cellular processes: ATP production, motility and active membrane transport. The two are interlinked, ordinarily PMF is used to synthesise ATP, but ATP can drive the production of PMF as well \citep{Keis2006}.

As early as 1791 Luigi Galvani proposed that life processes generate electricity \citep{Galvani1791,Green1953}. However, it took more than a century for Hugo Fricke to measure the capacitance of biological membrane \citep{Fricke1923} and for Peter Mitchell to explain that PMF is an electrochemical gradient of protons across the membrane that powers the production of ATP \citep{Mitchell1961}. PMF consists of the two components: pH difference between cytoplasm and the external environment ($\Delta pH=pH_{in}-pH_{out}$), and the electrical potential across the membrane ($V_{m}$, we note that the build up of charge occurs at $\sim$~nm-thin layer close to the biological membrane \citep{Nelson2003}).

\begin{equation}\label{eq:PMF}
PMF=V_{m}-\frac{2.303kT}{e} \Delta pH
\end{equation}

\noindent
where $k$ is the Boltzmann constant, $T$ is the temperature and $e$ is the elementary charge.

Since life generates electricity used to power its processes and cell membrane acts as a capacitor, it is reasonable to represent the rest of the cell components with an electrical circuit analogy \citep{VanRotterdam2002a,Walter2007}, Fig.~\ref{fig 1}A. Then, proton fluxes are currents, oxidative or substrate-level phosphorylation can be considered as an imperfect battery with non-zero internal resistance, and the membrane resistance and capacitance are connected in parallel. If external pH equals that of the bacterial cytoplasm, and for \textit{Escherichia coli} the latter is known \citep{Slonczewski1981,Zilberstein1984,Wilks2007}, $V_{m}$ in the circuit equals the PMF. 

The circuit analogy in Fig.~\ref{fig 1}A gives a mathematical framework that helps us understand cellular free energy maintenance in a range of different conditions. For example, we can predict changes in $V_{m}$ when circuit parameters change: a battery depends on the available carbon source and internal resistance R\textsubscript{i} increases in presence of electron transport chain inhibitors (such as sodium azide \citep{Noumi1987}). Furthermore, if we could measure $V_{m}$ with an equivalent of a "voltmeter" we could predict the mechanism and dynamics of the damage as the cells are exposed to various external stresses, as well as obtain functional dependence between affected circuit parameters and the amplitude of the stress. 

Here we report the use of bacterial flagellar motor as such a "voltmeter" and reveal the mechanisms of damage caused by chosen stresses. We confirm the behaviour of a known ionophore (indole) \citep{Chimerel2012}, discover that butanol is an ionophore, and quantitatively describe the nature of damage caused by the light of shorter wavelengths. Our approach of combining high-precision PMF ($V_m$) measurements and the "electrical circuit interpretation" of the cell serves as a powerful tool needed for quantitative bacterial electrophysiology.

\section*{Materials and Methods}
\subsection*{\textit{E. coli} strains}
\textit{E. coli} EK07 strain is constructed as described in \textit{Supplementary Materials}. Highly motile \textit{E. coli} strain MG1655 with an insertion sequence element in the \textit{flhD} operon \citep{Barker2004} is modified to have \textit{fliC} gene replaced by \textit{ fliC\textsuperscript{sticky}} \citep{Kuwajima1988}, which produces flagellar filaments that stick to glass or polystyrene surfaces. Additionally, \textit{pHluorin} \citep{Miesenbock1998,Morimoto2011} gene under strong constitutive \textit{Vibrio harveyi} cytochrome C oxidase promoter \citep{Pilizota2012} is placed onto \textit{att}Tn7 site of the chromosome. All the chromosomal alterations are generated using plasmid mediated gene replacement technique \citep{Link1997}.

\subsection*{\textit{E. coli} growth and media}
EK07 cells are grown in Lysogeny broth (LB: 10 g tryptone, 5 g yeast extract, 10 g NaCl per 1 L) after diluting from the overnight culture as 1:2000, at 37\degree C with shaking (220 rpm) to OD=2.0 (Spectronic 200E Spectrophotometer, Thermo Scientific, UK). We found that the yield of the single motor experiment, which is defined as the number of spinning beads in the field of view and likely corresponding to the flagellar motor expression level, was maximised at this OD in agreement with \citep{Amsler1993}. Growth curves of the EK07 and the parent MG1655 strain are given in SI Fig. \ref{SI growth curves}. After growth cells
are washed (3 times by centrifugation at 8000 g for 2~min) into MM9 (aqueous solution of 50~mM Na\textsubscript{2}HPO\textsubscript{4}, 25~mM NaH\textsubscript{2}PO\textsubscript{4}, 8.5~mM NaCl and 18.7 mM NH\textsubscript{4}Cl with added 0.1~mM CaCl\textsubscript{2}, 1 mM~KCl, 2~mM MgSO\textsubscript{4} and 0.3\% D-glucose) adjusted to pH=7.5 or PBS (aqueous solution of 154 mM NaCl, 5 mM Na\textsubscript{2}HPO\textsubscript{4} and 1.5 mM KH\textsubscript{2}PO\textsubscript{4}) adjusted to pH=7.5. Indole treatment is performed in MM9 and butanol and photodamage experiments in MM9 and PBS. 

\subsection*{Microscope slides preparation}
To shorten flagella, cells are "sheared" as described previously \citep{Bai2010,Rosko2017a} and washed as above. 
For butanol and indole treatment tunnel-slides are prepared as before \citep{Rosko2017a}, see also SI Fig. \ref{SI setup}A). For photodamage experiments flow-cells are manufactured by drilling (AcerDent, UK) two 1.8~mm holes on opposite ends of the microscope slide and attaching Tygon\textsuperscript{®} Microbore tubing (Saint Gobain Performance Plastics, France). The flow-cell is then created by attaching gene frame (Fisher Scientific Ltd, USA) to the slide and covering it with a cover glass (SI Fig. \ref{SI setup}B). Surface of the cover slide is coated with 0.1\% poly-L-lysine (PLL) by flushing PLL through the flow-cell/tunnel-slide for $\sim$10~s followed by washing it out with the excessive volume of growth medium. Sheared and washed cells are then loaded into the flow-cell/tunnel-slide and incubated for 10~min to allow attachment. Excessive cells are washed out with the growth medium. Subsequently, 0.5~$\mu$m in diameter polystyrene beads (Polysciences, Inc, USA) are added to the flow-cell/tunnel-slide and incubated for 10~min with consequent washing out of the non-attached beads. 

\subsection*{Microscopy and data collection}
Back-focal-plane interferometry \citep{Denk1990,Svoboda1993} is performed as previously described \citep{Rosko2017a}. Briefly, heavily attenuated optical trap (855~nm laser) is used to detect the rotation of a polystyrene bead attached to a truncated flagellar filament (Fig. \ref{fig 1}B). Time course of the bead rotation is recorded with the position-sensitive detector (PSD Model 2931, New Focus, USA) at 10~kHz, and a 2.5~kHz cutoff anti-aliasing filter applied before processing (Fig. \ref{fig 1}B). Bead position (x,y) is calculated from photocurrents $I_1 - I_4$ as $(I_1+I_2-(I_3+I_4))/(I_1+I_3+I_2+I_4) =2x/L$ and $(I_1+I_3-(I_2+I_4))/(I_1+I_3+I_2+I_4) =2y/L$, where L is the PSD detector side length.

Fluorescent images of pH sensitive pHluorin are taken in the same custom-built microscope with iXon Ultra EMCCD camera (Andor, UK). OptoLED Dual (Cairn Research Ltd, UK) independently driving two LEDs is used for the illumination. Narrow spectrum UV LED is used for excitation at 395~nm and Neutral White LED with ET470/40x filter (Chroma Technology, USA) for 475~nm excitation. Emission is taken at 520~nm using ET525/40x filter (Chroma Technology, USA). Exposure time is fixed at 50~ms for butanol and indole treatment experiments and varies from 10 to 200~ms for photodamage experiments. 

\subsection*{Applying stresses}
Butanol (1-Butanol for molecular biology, $\geqslant$99\%, Sigma-Aldrich, USA) and indole (Indole, analytical standard, Sigma-Aldrich, USA) treatment is performed as follows: after recording the motor speed for 2~min, 20~$\mu$l of MM9 (or PBS)  supplemented with a given concentration of butanol or indole is flushed into the tunnel-slide. Flush is done by placing a droplet of liquid on one, and collecting it with a piece of tissue paper on the other side of the tunnel \citep{Buda2016}. Duration of the flush is no longer than 10~s. 10~$\mu$l droplets of shocking solution are then placed on both side of the tunnel to minimise evaporation. The shock motor speed is recorded for 10~min, followed by a flush back into MM9 (or PBS) medium. Postshock speed is recorded for 5~min. The motor speed recording is uninterrupted for the duration of the experiment (total of 17~min). For pH control experiments fluorescent images are taken every 90 seconds. Control flushes with media containing no indole/butanol are shown in SI Fig. \ref{SI flush control}.

Photodamage experiments are performed as follows: using the flow-cell MM9 or PBS is constantly supplied at 10~$\mu$l/min rate with a syringe pump (Fusion 400, Chemyx Inc., USA). Cells are sequentially exposed to the light of $\lambda$=395~nm and 475~nm. Speed recording starts simultaneously with the light exposure. The camera exposure time ($t_{exp,cam}$) and sampling rate are controlled with a custom written LabView program. $t_{exp,cam}$ are set the same for both wavelengths, however hardware adds a different delay, thus effective light exposure times are $t_{exp,light}=$225~ms+$t_{exp,cam}$ for 475~nm and 55~ms+$t_{exp,cam}$ for 395~nm. We record $t_{exp,light}$ and sampling rate throughout the experiment to calculate the effective light power ($P_{eff}$) as the total energy delivered, divided by the total length of the individual motor speed recording. Total energy delivered is estimated by measuring the illumination power in the sample plane multiplied by the total time of light exposure and divided by the illumination area. We measured the illumination area by photobleaching part of the slide and measuring the diameter of the bleached region (d $\approx$ 220~$\mu$m). Control speed traces with no light exposure are shown in SI Fig.~\ref{SI control}.

\subsection*{Data analysis}
A flat-top window discrete Fourier transform (window size=16384 data points with $dt=$0.01~s) is applied to $x$ and $y$ coordinates of bead position to obtain a time series motor speed record. This speed records we refer to as raw speed traces (Fig. \ref{fig 1}C, \ref{fig2}A, \ref{fig3}A, \ref{fig:4}A, SI Fig. \ref{SI flush control}, \ref{SI control}, \ref{SI multiflush}, \ref{SI recovery}). Raw traces are further processed as follows: (a) absolute values are taken, (b) values below 10~Hz are removed and 50~Hz AC frequency values disregarded, (c) remaining data points are median filtered with 201 points moving window. To calculate mean speeds we apply a 10~s moving window on the speed traces processed as above.
In addition to above, photodamage traces are normalised. First, 30~s of the trace is split into 60 windows containing 50 points each. The mean of maximum values found within each window is calculated and considered the initial speed value, by which the rest of the trace is normalised. Each normalised trace is fitted with a single parameter exponential: $y=e\textsuperscript{-$\alpha$x}$. For Fig. \ref{fig2}D and \ref{fig3}D hyperbolic function fitted is $y=\frac{1}{Kx+1}$ and quadratic hyperbolic $y=\frac{1}{Kx^2+1}$, where K is a fitting parameter. All fittings are performed in Python (SciPy module, curve fit optimization) with maximum number of calls to the optimization function taken as 20~000.

\section*{Results}
\label{sec:results}

\subsection*{PMF measurements via flagellar motor speed can be used to analyse stress-induced damage.}
\begin{figure}[h]
\centerline{\includegraphics[width=1\linewidth]{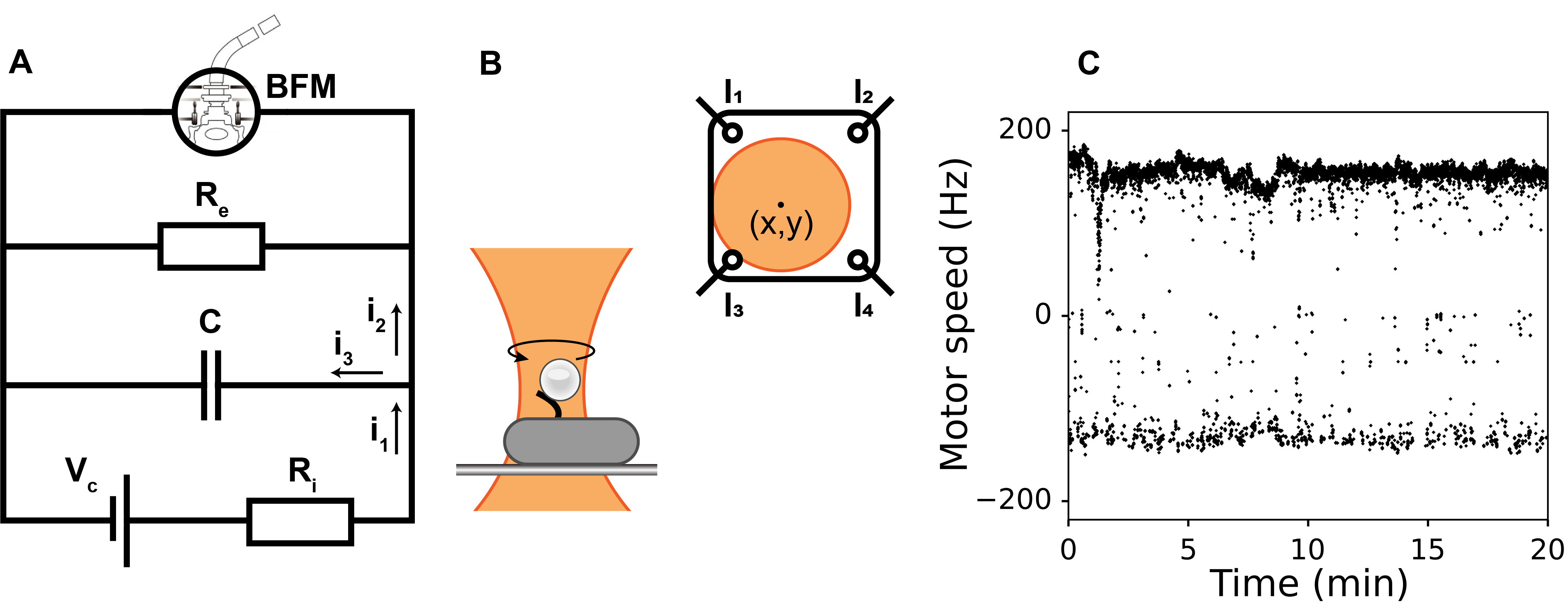}}  
\centering
\captionsetup{justification=justified}
\caption{ (\textbf{A}) Electric circuit equivalent of an\textit{ E. coli} cell. Oxidative (or substrate-level) phosphorylation is shown as a battery $V_{c}$ with an internal resistance R\textsubscript{i}, the membrane with capacitance C and resistance R\textsubscript{e}, and $i_1$ to $i_3$ are the currents. Bacterial flagellar motor (BFM) is shown as a "voltmeter" that measures membrane potential, $V_m$. (\textbf{B}) Schematic of the "bead-assay" and back-focal-plane interferometry. A cell is attached to a cover glass with a truncated flagellar filament made "sticky" to polystyrene beads. The bead is brought into a heavily attenuated optical trap and its position measured with position sensitive detector (see \textit{Materials and Methods}). (\textbf{C}) An example of raw motor speed trace recorded with back-focal-plane interferometry. Positive frequencies correspond to counter-clockwise and negative to the clockwise rotation of the flagellar motor \citep{Bai2010}. In the subsequent figures we show absolute values of the rotational speeds.}
\label{fig 1} 
\end{figure}

The electric circuit analogy (Fig.~\ref{fig 1}A) gives a mathematical framework needed to understand cellular free energy maintenance in a range of different conditions. For example, under given external stress it allows us to discern the affected component of the cell and predict the mechanism of damage caused by the stress in the following manner. Membrane capacitance is set by the geometry of the lipid bilayer and unlikely to be altered on shorter time scales. $V_c$ is the theoretical maximum potential a cell can generate in  a given environment and from a given internalised (carbon) source. Stress can affect $V_c$ only by damaging specific carbon transporters and, thus, is media-dependent. Furthermore, in starvation buffer where \textit{E. coli} uses internal carbon sources \citep{Nystrom1998} $V_c$ will not be changed by the stress. $R_i$ defines the inefficiency of the catabolism, comprising the drop from $V_c$ as a specific carbon source gets metabolised via a large number of catabolic enzymes. These enzymes are at least partially carbon source specific, thus the stress that targets $R_i$ will be media-dependent. Finally, while the $R_e$ value is growth media-dependent, the membrane targeting stresses that influence $R_e$ will be media-independent. 

Once we pin-down the affected component, we employ Kirchoff's laws to  express it as a function of stress-induced membrane potential change ($V_m/V_{m,0}$), which we measure using bacterial flagellar motor (BFM) as a "voltmeter"(Fig.~\ref{fig 1}A). 
BFM is a nano-machine that enables bacterial swimming \citep{Sowa2008} via PMF powered rotation \citep{Manson1980,Matsuura1977,Meister1987,Fung1995}. The motor speed ($\omega$), usually reaching couple of hundred Hz \citep{Lowe1987}, varies linearly with PMF \citep{Fung1995,Gabel2003}. While BFM can be actively slowed down, e.g. when cell enter stationary phase \citep{Amsler1993}, on shorter time scales the linearity between the motor speed and PMF allows us to use the motor speed as a PMF indicator, and when $pH_{cytoplasm}=pH_{external}$ as a $V_{m}$ indicator as well. Here we consider only the situation where $\Delta pH\approx0$, which we set by adjusting the external pH to known internal pH of \textit{E. coli} \citep{Slonczewski1981}, and in the rest of the text use PMF and $V_m$ interchangeably. In addition, EK07 strain we constructed (see \textit{Materials and Methods}) carries a chromosomal copy of the gene encoding pHluorin protein, which we use to check that our expectation is correct, i.e. cytoplasmic pH during the experiments stays constant and at the level of external pH (SI Fig. \ref{SI pH}). We thus have: 

\begin{subequations}\label{first:motorspeed}
\begin{equation} \label{motorspeed:a}
\omega=\xi*PMF=\xi*V_{m}
\end{equation}

\begin{equation} \label{motorspeed:b}
\frac{\omega}{\omega_0}=\frac{PMF}{PMF_0}=\frac{V_{m}}{V_{m,0}}=f(S,t)
\end{equation}
\end{subequations}

\noindent
where we assumed that $\omega$ changes as a function of stress amplitude and time $f(S,t)$, $\xi$ is a constant and index 0 denotes the variable value prior to stress. We measure $\omega$ using back-focal-plane interferometry \citep{Svoboda1993} and a polystyrene bead attached to a short filament stub (see \textit{Materials and Methods} and Fig.~~\ref{fig 1}B) \citep{Bai2010}. An example trace of BFM speed is given in Fig.~~\ref{fig 1}C. Using equation \eqref{motorspeed:b} and the circuit analogy we can express each circuit component as a function of stress. To do so, we simplify the electric circuit by estimating the $RC$ constant of the cell membrane. Capacitance and resistance of the bacterial membrane have been reported as C$\sim$1~$\mu$F/$cm^2$ \citep{Fricke1956,Hodgkin1952} and R$\sim$10-1000 $Ohm*cm^2$ \citep{Miyamoto1967,Chimerel2012}, which gives $RC$ in the range of $10^{-5}$ to $10^{-3}$~s. Thus, the current through the capacitor is zero prior to the stress application (when the system is in steady-state), as well as post stress application when $t> 1~ms$ (i.e. on the time scales of our experiment).
Next we consider $\Delta$G of NADH oxidation only, and compute that respiratory chain can produce $V_{c}$ $\sim$-360~mV \citep{Walter2007}. Yet, physiological value of the membrane potential of respiring bacteria is approximately equal to -160~mV \citep{Tran1998}, indicating that roughly half of the membrane potential drops at the internal resistance, i.e. $R_{i,0}\approx R_{e,0}$. Taking the two simplifications into account we arrive to (see Fig.~\ref{fig 1}A and \textit{Supplementary Material} for detailed deduction of equations):

\begin{subequations}\label{first:fcomp}
\begin{equation}
\frac{V_c}{V_{c,0}} =f(S,t)
\label{fcomp:c}
\end{equation}

\begin{equation}
\frac{R_i}{R_{i,0}} =\frac{2}{f(S,t)} -1 
\label{fcomp:a}
\end{equation}

\begin{equation}
\frac{R_e}{R_{e,0}} =\frac{f(S,t)}{2-f(S,t)}  
\label{fcomp:b}
\end{equation}
\end{subequations}

\subsection*{PMF dynamics analysis confirms indole is an ionophore}
We test the proposed circuit analogy and applicability of the BFM speed as the voltmeter by applying a known membrane stress. We choose a cell signaling molecule indole that at millimolar concentrations forms a dimer and acts as an ionophore \citep{Chimerel2012}. Ionophores are molecules that carry ions across the lipid bilayer, thus we expect the membrane resistance to decrease (ion conductance increases) when indole is present in the medium. Furthermore, we expect to recover previously demonstrated parabolic dependence of membrane conductance on indole concentration \citep{Chimerel2012}.

Fig.~2A shows examples of individual motor speed recordings prior, during and post treatment with a given concentration of indole. Motor speed drops immediately with the addition of indole, and stays at approximately the same level until indole is removed, at which point it recovers to the initial level. The speed change caused by indole is faster than 10~ms (our experimental resolution), confirming the estimate of membrane RC constant, and justifying the assumption that the current through the capacitance in Fig. \ref{fig 1}A circuit is negligible.

\begin{figure}[h!]
\centerline{\includegraphics[width=0.8\linewidth]{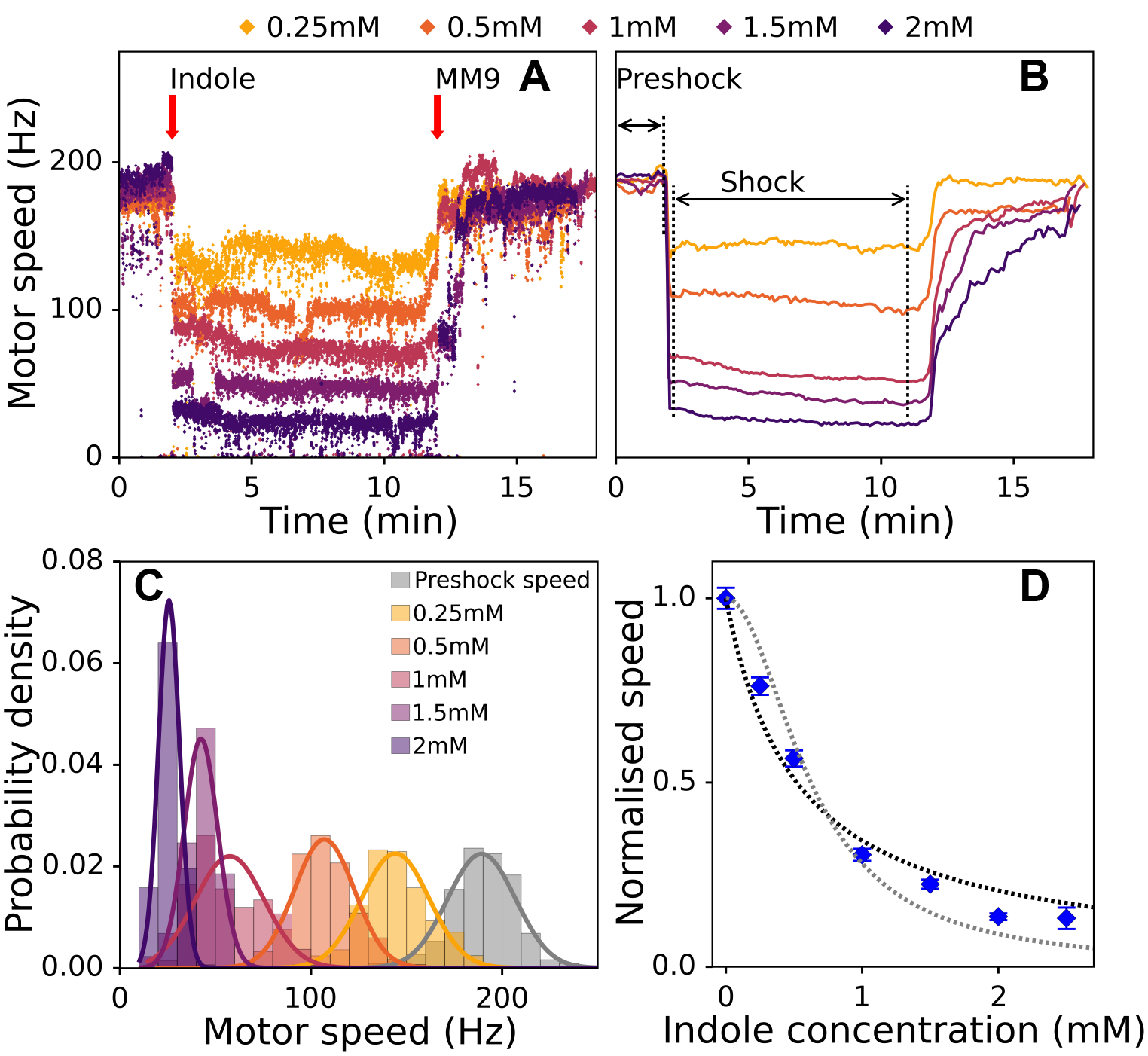}}
\centering
\captionsetup{justification=justified}
\caption{BFM speed drops rapidly and increasingly with increasing indole concentration (\textbf{A}) Examples of raw motor speed traces at 5 different indole concentrations. Indole is delivered into the tunnel-slide 2~min after the recording commences, and removed after 12~min. (\textbf{B}) Mean speeds of n$\geqslant$20 motor speeds for each indole concentration are shown against time. Each motor recording is performed on a different cell, thus the number of motors corresponds to the number of different individual cells. Preshock speed is calculated for time interval between 0 and 110~s (indicated in the figure). Shock speed is calculated from the 130 to 660~s of the motor recording. Preshock and shock intervals were chosen to exclude the duration of the flush. Standard errors are given, but not visible (for standard deviations see SI Fig. \ref{SI std}A). (\textbf{C}) Probability density of motor speeds for each indole concentration. Experimental data is fitted with a Gaussian probability density function. (\textbf{D}) Normalised BMF speeds plotted against indole concentration. Error bars represent standard error of the mean, and dotted lines show hyperbolic (black) and quadratic hyperbolic (grey) fit ($R^2 = 0.97$ and $R^2 = 0.95$ respectively).}
\label{fig2} 
\end{figure}

To confirm the dependence of the membrane resistance on indole we find the relative change in motor speed at a given stress concentration. Fig. \ref{fig2}B shows the mean speed traces for different indole concentrations (see \textit{Materials and Methods} for mean speed calculation) and in Fig. \ref{fig2}C we plot the probability densities of preshock and shock speeds. From the Gaussian fits to preshock and shock speed distributions we obtain mean shock speeds for a given indole concentration, and plot them normalised to the preshock speed, Fig. \ref{fig2}D. We fit the normalised speeds with hyperbolic or quadratic hyperbolic function (see \textit{Materials and Methods}, both of which yield good quality fits with $R^2$ higher than 0.90). The concentrations of indole at which the quadratic dependence becomes particularly obvious are higher than 2.5~mM \citep{Chimerel2012}, where we use 0-2.5~mM range. Therefore, our result confirms the accuracy of our proposed approach.

\subsection*{Butanol acts as an ionophore, changing membrane conductance linearly with concentration.}
\begin{figure}[h!]
\centerline{\includegraphics[width=0.8\linewidth]{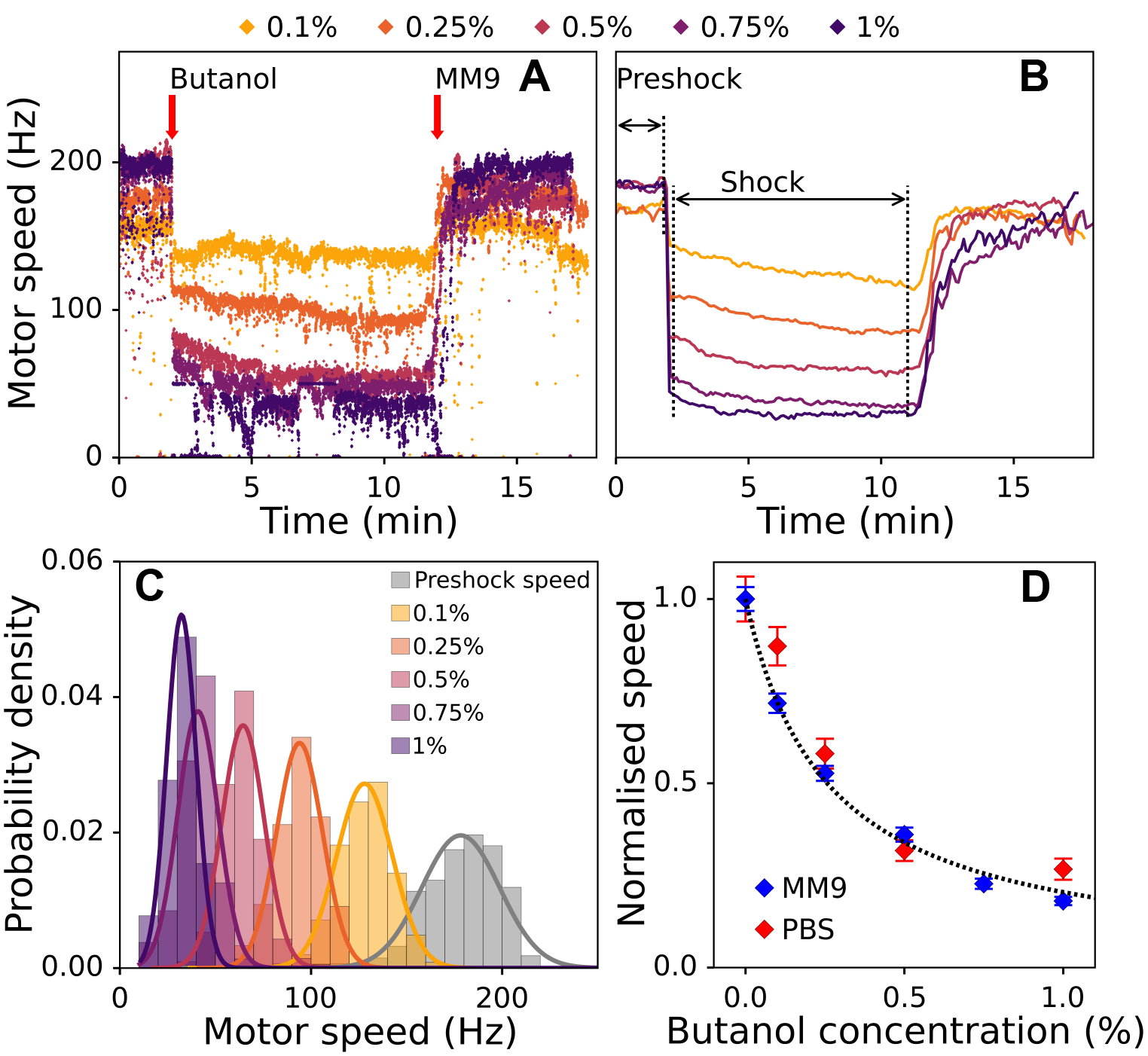}}
\centering
\captionsetup{justification=justified}
\caption{BFM speed drops sharply and reversibly after butanol treatment. (\textbf{A}) Examples of raw BFM speed traces for 5 different butanol concentrations. Butanol is delivered 2~min into the recording and removed after 12~min. (\textbf{B}) Mean speeds of n$\geqslant$20 cells per different butanol concentrations are plotted against time. Preshock and shock speeds are calculated in the 0 to 110~s, and 130 to 660~s time interval, respectively. Standard errors of the mean are given, but not visible. Standard deviations of the same traces are given in SI Fig. \ref{SI std}B. (\textbf{C}) Probability densities of shock speed for each butanol concentration and the preshock speed. (\textbf{D}) Shock speeds obtained from the distributions are normalised by the preshock speed and plotted against butanol concentration. Blue diamonds show cells in MM9 media and red diamonds cells in PBS. Error bars represent standard error of the mean. Hyperbolic fit is given as a black dotted line ($R^2$ = 0.96).}
\label{fig3} 
\end{figure}

To determine the mechanism of action of an unknown stress we choose butanol. Previous work indicates that butanol interacts with cell membrane and weakens it, but the exact mechanism of cell damage is unknown \citep{Fletcher2016}. We perform the BFM speed measurements in \textit{E. coli} cells treated with butanol. The experimental protocol of butanol delivery is the same as for indole. Fig. \ref{fig3}A and \ref{fig3}B show examples of raw traces and mean speed traces prior, during and post butanol shock in MM9. Immediately upon butanol stress motor speed drops, and upon butanol removal it recovers to the initial value, Fig. \ref{fig3}A. Motor speed distributions at a given butanol concentrations remain narrow, and we fit them with Gaussian curves (Fig. \ref{fig3}C). Fig. \ref{fig3}D shows normalised motor speeds, calculated as mean values of the distributions given in Fig. \ref{fig3}C, and plotted against butanol concentration for both MM9 media and PBS.

The relative speed drop observed in the presence of butanol is media independent, and alike that observed for indole.
The finding suggesting that, on the time scale of our experiment, butanol cause non-permanent membrane damage and acts as an ionophore. The normalised motor speed dependence on butanol concentration is hyperbolic, and we obtain equation \eqref{fcomp:b} for membrane resistance:

\begin{equation} \label{Rebut} 
R_e=\frac{R_{e,0}}{7.8*c_{but}+1}
\end{equation}
\noindent
where $c_{but}$ is a butanol concentration in percents (\%) and 7.8 is a value of constant $K$ obtained from the hyperbolic fit (see \textit{Materials and Methods}). We observe the speed restoration after butanol removal even after multiple treatments of the same cell. SI Fig. \ref{SI multiflush} shows several consecutive butanol stresses each lasting 60~s (A) or 30~s (B), where after each treatment motor speed is fully restored.

\subsection*{Photodamage increases membrane conductance that scales with the light power. }

As an example of a complex stress we next choose to characterise light induced damage. While previous reports indicate that light causes wavelength dependent damage to bacterial cells \citep{Ashkin1987,Neuman1999}, they also suggest that the nature of damage is complex. Most likely the cause of the damage is formation of reactive oxygen species (ROS) \citep{DeJager2017,Lockwood2005}, which have been shown to perturb multiple components of the cell: DNA, RNA, proteins and lipids \citep{Cabiscol2000,Zhao2014}. 
To apply light of a certain wavelength and intensity to bacterial cells we use a flow-cell (see \textit{Materials and Methods}). During the light exposure cells are continuously supplied with fresh media at 10~$\mu$l/min flow rate. We apply the light of 395~nm and 475~nm wavelengths as the choice allows us to simultaneously measure internal pH of bacteria.

\begin{figure}[h!]
\centerline{\includegraphics[width=1\linewidth]{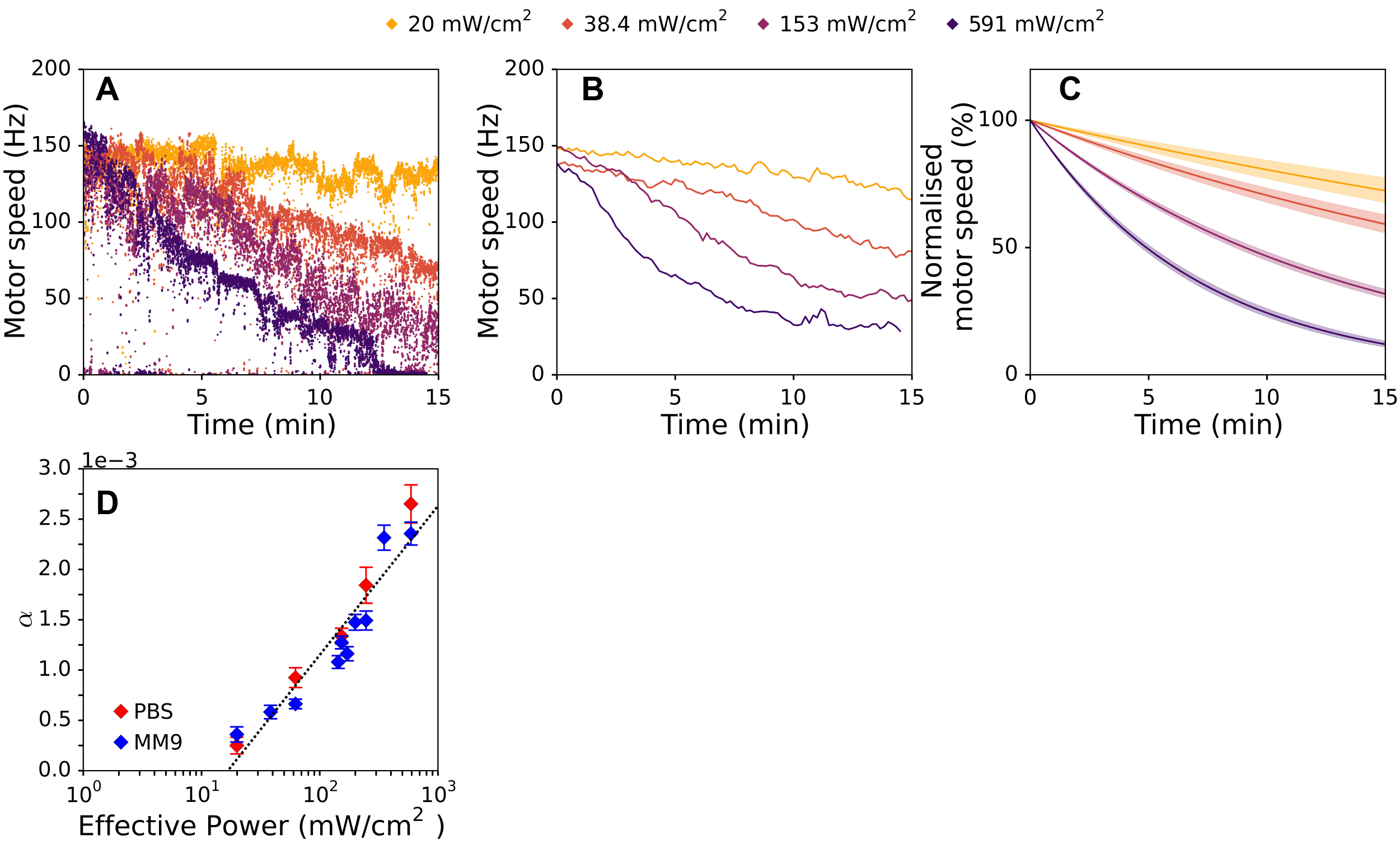}}
\centering
\captionsetup{justification=justified}
\caption{Rate of the motor speed decay increases with the light power. (\textbf{A}) Examples of raw traces at four different effective powers, $P_{eff}$ = 20, 38.4, 153 or 591 mW/$cm^2$. (\textbf{B}) Mean BFM speed at different illumination powers (21 to 34 cells are recorded per condition). (\textbf{C}) Averaged exponential fits for different illumination powers with standard error. Each individual motor trace is fitted with an exponential function and the mean of fitting parameter $\alpha$ is calculated for each $P_{eff}$. (\textbf{D}) Exponential fit coefficient $\alpha$ is plotted against illumination power. Blue diamonds show cells in MM9 media and red diamonds cells in PBS. Error bars represent standard error and dotted line the logarithmic fit ($R^2$ = 0.906). The total number of cells in MM9 is 277 and in PBS 116.}

\label{fig:4} 
\end{figure}

Fig. \ref{fig:4}A shows example BFM speed traces during exposure to light of different effective powers ($P_{eff}$) delivered to the cells. $P_{eff}$ is calculated as the total energy delivered divided by the total time the light is on (see \textit{Materials and Methods}). Fig. \ref{fig:4}A shows that BFM speed gradually decreases in time during exposure to light and that the decrease rate scales with the $P_{eff}$, also visible in Fig. \ref{fig:4}B showing mean BFM speed traces for the same four effective powers. 

To identify the functional dependence of the speed decrease rate on $P_{eff}$ we fit individual normalised traces with the simple exponential function: $\omega/\omega_{0} = e^{-\alpha t}$, with the single fitting parameter $\alpha$. Mean of the fits with standard errors at corresponding four different powers are shown in Fig. \ref{fig:4}C, and Fig. \ref{fig:4}D shows fit coefficient $\alpha$ plotted against the light power for both MM9 medium and PBS. The effect of light on $V_{m}$ is present in PBS and of same functional dependence, thus on the time scales of our experiment light affects primarily the membrane resistance, $R_e$. Together with the fact that the speed decrease rates stay the same at a given $P_{eff}$, the finding suggests that on the time scale of our experiment there is no active membrane repair. We further confirm this by measuring the motor speed after we expose the cells to light for shorter periods of time. SI Fig. \ref{SI recovery} shows that when the illumination ceases after 5 or 15~min the (decreased) BFM speed remains the same with no visible recovery.
We also check that light damage is not enhanced by the presence of the fluorescent protein (pHluorin) in the cytoplasm, SI Fig. \ref{SI fluorophore}.

Fig. \ref{fig:4}D enables us to determine functional relationship between effective power and $\alpha$, which increases as a logarithm of the normalised $P_{eff}$, i.e. $P_{eff,norm}=P_{eff}/(mW*cm^{-2})$. Thus, for our initial exponential fit we obtain:

\begin{equation}
\omega = \omega_0*e^{-(a\ln{P_{eff,norm}}+b)t}    \label{light log}
\end{equation}

\noindent
where $a$ and $b$ are wavelength specific parameters, $a$ = 0.00064~$s^{-1}$ and $b$ = -0.00181~$s^{-1}$ and equation \eqref{light log} holds for $P_{eff}>P_{eff,0}$.
 
 The minimum power required for the damage to occur is defined as $P_{eff,0}=e^{-\frac{b}{a}}~mW/cm^{2}$, and for 395~nm and 475~nm this is $\sim$ 17~mW/$cm^{2}$. Re-writing the equation \eqref{light log} in terms of $P_{eff,0}$ we get:

\begin{equation}
\omega = \omega_0 \left(\frac{P_{eff,0}}{P_{eff}}\right)^{at}    \label{light omega}
\end{equation}

\noindent
Finally, applying \eqref{light omega} to equation \eqref{fcomp:b} we derive $R_e$ functional dependence on the effective power:

\begin{equation}
R_e =  \frac{R_{e,0}}{2\left(\frac{P_{eff}}{P_{eff,0}}\right)^{at}-1},    \label{light resistance}
\end{equation}
 
\noindent 
where $a$ is the fit coefficient in Fig. \ref{fig:4}D.

\section*{Discussion}
Arguably, one of the defining features of life is its ability to avoid thermodynamic equilibrium (death) by achieving a steady state supply of free energy. Chemiosmotic theory explained that the production of life's energy currency, the ATP molecule, proceeds via the generation of trans-membrane electrochemical potential. The ability to measure and control voltage and current across the cellular membrane with the patch-clamp technique had far reaching consequences for our understanding of cells such as neurones, where the electrical inputs govern signal transmission \citep{Hodgkin1952}. In the cases of bacteria, and their small size, we are unable to gain the same level of control over these parameters \citep{Ruthe1985,Martinac1987}, despite the fact that the ability to do so would open a range of currently inaccessible questions that are at the basis of bacterial free energy maintenance, and consequently, survival.

Here we demonstrate the use of BFM as a fast voltmeter, enabling quantitative, \textit{in vivo} studies of electrochemical properties of the bacterial membrane. Alternative methods for measuring $V_{m}$ in \textit{E. coli} rely on fluorescent readout \citep{Ehrenberg1988,Prindle2015, Kralj2011}. However, Nernstian dyes \citep{Ehrenberg1988,Prindle2015} sometimes fail to penetrate \textit{E. coli's} membrane \citep{Lo2007}, can be a substrate for the outer membrane efflux system TolC \citep{Mancini2018} and in external conditions where they do equilibrate across the membrane, they do so on the time scales of minutes \citep{Lo2007,TeWinkel2016}. Voltage sensitive membrane proteins that can be used in \textit{E. coli} require delivery of light of high power \citep{Kralj2011}. BFM, on the other hand, is native to \textit{E. coli} and expressed in a range of conditions \citep{Cremer2018}. Speed measurements via back-focal-plane interferometry or fast cameras do not rely on fluorescent illumination and offer high time resolution (up to ~0.5~ms \citep{Pilizota2007}). We choose to work with cells grown into late exponential phase in LB to maximise our experimental yield. The approach is, however, more widely applicable as BFM is expressed in a range of other conditions (with the exception of late stationary phase cells \citep{Amsler1993,Cremer2018}), where we expect the BFM bead-assay yield to somewhat vary with the condition. It is possible that cells grown to early or mid exponential phase, or cells grown to steady state in different growth media, will have different electrochemical properties, which can be measured with our approach in the future. 

We choose to work in the conditions that satisfy $\Delta$pH$\approx$0, and thus $V_{m}$ is the only contribution to the PMF. However, BFM speed measurements can be extended to conditions where $\Delta$pH contribution to the PMF is not negligible, $V_{m}$ in this case will be calculated from equation \eqref{eq:PMF}. Extending the use of BFM as the voltmeter for long term measurements (into hours and days) is possible. We note that on longer time scales motor can be actively slowed down via YcgR protein \citep{Boehm2010,Paul2010}, and such long term measurements would likely require YcgR deletion background.
  
We base the use of BFM as the cell's voltmeter on the proportionality between motor speed and PMF, measured first more than 20 years ago \citep{Fung1995,Gabel2003}. Recent experiments show that BFM also exhibits mechanosensing \citep{Lele2013,Tipping2013}, where stator unit incorporation depends on the motor torque. These recent findings indicate an intriguing control mechanism, where mechanosensing and the ion flux combined result in the characteristic proportional relationship between the BFM speed and PMF. It will be interesting to fully ascertain the exact molecular mechanism behind the PMF-motor speed relationship, and we think the ability to fine-control the PMF loss can contribute to that understanding.

Using the electric circuit analogy for the membrane fluxes, and BFM as the cell's "voltmeter" we demonstrate the effect of three different stresses on the cell's membrane conductance. For the known stress, indole, we confirm it acts as an ionophore. For the first unknown stress we applied, butanol, we show its presence decreases membrane resistance, inversely proportional to the butanol concentration. Thus, we conclude that, in the concentration range we tested and on the 15~min time scale, butanol behaves as an ionophore in a manner similar to indole or CCCP \citep{Chimerel2012}. With analysis alike we presented, butanol action can be characterised further, e. g. defining the minimum concentration and incubation time required for the effect to become irreversible. 
For our last stress, light of short wavelengths, we show that it affects membrane resistance and functionally describe the damage in relation to time and $P_{eff}$. Light-induced changes in membrane permeability have been reported in artificial planar lipid bilayer systems and liposomes in the presence of photosensitisers \citep{McRae1985,Pashkovskaya2010,Kotova2011,Wong-Ekkabut2007}. The most likely cause of such changes is ROS induced chain-reaction lipid peroxidation \citep{Girotti1985,Girotti1990,Halliwell1993,Heck2003a,Lavi2010}. Presence of peroxidised lipids can change bilayer physical and electrical properties \citep{Dobretsov1977,Richter1987,Birben2012}, e.g. it has been suggested that it induces formation of hydrophobic pre-pores and their later transformation into hydrophilic pores permeable to ions \citep{Kotova2011,Wong-Ekkabut2007}. 
Based on the previous work, and our real time, \textit{in vivo} measurements we propose the following model for the complex nature of the light-induced membrane damage. Exposure to light causes the formation of ROS that induce lipid peroxidation, and thus alter the electric properties of the membrane. In particular, its permeability to ions due to the formation of hydrophilic pores. In contrast to the ionophores that carry ions across the membrane without causing membrane damage, the drop in $V_m$ we observe under light proceeds as a result of slower, multi-step formation of lipid pores that require active repair to be mitigated. Therefore, we do not see any fast recovery after illumination ceases (SI Fig. \ref{SI recovery}), and the chain-reaction nature of the process results in the exponential-like decay of membrane potential.

Living cells have built-in mechanisms of coping with oxidative stress, for example SoxRS/OxyR regulons containing multiple antioxidant-encoding genes, such as \textit{sodA} (manganese superoxide dismutase) or \textit{katG} (hydroperoxidase I) \citep{Storz1999,Birben2012}. The existence of defence mechanisms explains the occurrence of the minimum power required to cause the damage. Less power, even if it causes ROS formation, will not damage the cells that cope using internal protection enzymes. The value of the minimal damage-causing power we measured can indicate the abundance of internal protective resources available to the cell, as well as define the power range for fluorescence imaging that should be used to ensures no (unaccounted for) damage is inflicted to the cells by the exposure to light.

Future applications of our approach include, but are not limited to, studying other damage mechanisms and characterising unknown bacterial membrane properties, e. g. overall resistance in different growth conditions. Lastly, based on our measurements we suggest the use of light for delivery of small molecules, such as antimicrobial peptides or fluorescent dyes, which otherwise fail to penetrate \textit{E. coli's} membrane \citep{Lo2007}.
 
\subsection*{Author Contributions}
EK, CJL and TP designed research. EK performed research and analysed data. EK, CJL and TP interpreted results and wrote the paper. 

\subsection*{Acknowledgements}
We thank all the members of Pilizota and Lo laboratories, Zaki Leghtas, Jelena Baranovic, Bai Fan, Peter Swain, Ivan Maryshev, Ivan Erofeev, Calin Guet and Munehiro Assaly for useful discussions. EK was supported by the Global Research and Principal's Career Development PhD Scholarships. TP and CJL were supported by the Human Frontier Science Program Grant (RGP0041/2015), and CJL by the Ministry of Science and Technology, Republic of China (MOST-106-2112-M-008-023).

\bibliographystyle{apalike}

\bibliography{scibib}
\newpage
\section*{Supplementary materials}
\subsection*{Deduction of electric circuit model}

Let us consider the circuit from Fig. \ref{fig 1}A without yet applying simplifications discussed in the main text. When the system is in equilibrium current trough the capacitor $i_3=0$. Upon application of a given stress one of the three components of the circuit change ($R_i$, $V_c$ or $R_e$), and all three currents become non zero. Then, based on Kirchoff's laws we have:

\begin{subequations}
\begin{equation}
V_c=i_1 R_i + i_2 R_e  
\tag{S1a}
\label{first:Kirchoff}
\end{equation}

\begin{equation}
i_1 = i_2 + i_3
\tag{S1b}
\label{Kirchoff:a}
\end{equation}

\begin{equation}
i_2 R_e = \frac{Q}{C} 
\tag{S1c}
\label{Kirchoff:b}
\end{equation}
\end{subequations}

\noindent
We do not consider scenarios that change more than one circuit element at the same time. Let us first consider the case when  overall membrane resistance, $R_e$, changes, i.e. membrane has been damaged and $R_{i}$, $V_{c}$ and $C$ are kept fixed. Based on Ohm's law membrane voltage can be expressed as $V_m=i_2 R_e$. Thus, to get the functional dependence of $R_{e}$ on the stress amplitude we need to find $V_{m}/V_{m,0}$, equation \eqref{motorspeed:b} and \eqref{fcomp:b} in the main text. 

\setcounter{equation}{0}

\begin{equation}
\frac{R_e}{R_{e,0}}=\frac{V_m}{V_{m,0}} * \frac{i_{2,0}}{i_{2}}
\tag{S2}
\label{re_re0}
\end{equation}
\noindent
where $i_{2,0}$ is the current through $R_{e,0}$, i.e. before $R_e$ changed. We are now looking for expressions for $i_2$($R_e$,t) and $i_{2,0}$. We know that the current through the capacitor is $i_3=dQ/dt$ and thus from \eqref{Kirchoff:b} follows:

\begin{equation}
\frac{di_2}{dt}R_e +i_2 \frac{dR_e}{dt}=\frac{i_3}{C} 
\tag{S3}
\label{der_Kirchoff}
\end{equation}

\noindent
We express $i_3$ from \eqref{der_Kirchoff} and $i_1$ from \eqref{Kirchoff:a} and apply it to \eqref{first:Kirchoff}:

\begin{equation}
\frac{di_2}{dt} + i_2 \frac{R_i+R_e+C R_i \frac{dR_e}{dt}}{C R_i R_e} - \frac{V_c}{C R_i R_e} = 0
\tag{S4}
\label{i2}
\end{equation}

\noindent
When $i_3=0$, $i_{2,0}=i_{1,0}=i_0$ based on equation \eqref{Kirchoff:a} and by implementing to \eqref{first:Kirchoff} we get:

\begin{equation}
i_{0}=\frac{V_c}{R_i+R_{e,0}}
\tag{S5}
\label{i0}
\end{equation}
\noindent
because $V_{c}$ and $R_{i}$ do not change during the application of the stress. Using \eqref{motorspeed:b} in the main text and \eqref{i0} we can now express \eqref{re_re0} as:

\begin{equation}
\frac{R_e}{R_{e,0}}=f(S,t)* \frac{V_{c}}{R_{i}+R_{e,0}}*\frac{1}{i_2(R_e,t)},
\tag{S6}
\end{equation}

\noindent
where $i_2$($R_e$,t) is a solution of \eqref{i2}. Alternatively, if the stress-affected element is $R_i$ or $V_c$ equation \eqref{der_Kirchoff} and \eqref{i2} become:

\begin{equation}
\frac{di_2}{dt}R_e=\frac{i_3}{C} 
\tag{S7}
\end{equation}

\begin{equation}
\frac{di_2}{dt} + i_2 \frac{R_i+R_e}{C R_i R_e} - \frac{V_c}{C R_i R_e} = 0
\tag{S8}
\end{equation}

We now apply simplifications mentioned in the main text: $i_3$ is always zero ($RC$ is in the range of $10^{-5}$ to $10^{-3}$~s), and $R_{i,0}\approx R_{e,0}$. Taking the two simplifications into account we arrive to:

\begin{equation}
i=\frac{V_c}{R_i+R_{e}}
\tag{S9}
\label{i}
\end{equation}

\noindent
and
\begin{equation}
f(S,t)=\frac{V_m}{V_{m,0}}=\frac{V_c*R_e}{R_i+R_e}*\frac{R_{i,0}+R_{e,0}}{V_{c,0}*R_{e,0}}=\frac{V_c*R_e}{R_i+R_e}*\frac{2}{V_{c,0}}
\tag{S10}
\label{Vmration_simp}
\end{equation}

\noindent
Then, if $R_e$ is affected by the stress, $V_c=V_{c,0}$ and $R_i=R_{i,0}=R_{e,0}$

\begin{equation}
\frac{V_m}{V_{m,0}}=\frac{2*R_e}{R_e+R_{e,0}}=f(S,t)
\tag{S11}
\end{equation}

\noindent
If $R_i$ is affected by the stress, $V_c=V_{c,0}$ and $R_e=R_{e,0}=R_{i,0}$

\begin{equation}
\frac{V_m}{V_{m,0}}=\frac{2*R_{i,0}}{R_i+R_{i,0}}=f(S,t)
\tag{S12}
\end{equation}

\noindent
Finally, if $V_c$ is affected by the stress, $R_e=R_{e,0}=R_i=R_{i,0}$

\begin{equation}
\frac{V_m}{V_{m,0}}=\frac{V_c}{V_{c,0}}=f(S,t)
\tag{S13}
\end{equation}

\subsection*{Supplementary Methods:\textit{E. coli} strain construction}
All the chromosomal manipulations were performed using plasmid mediated gene replacement method (PMGR) described previously \citep{Link1997,Merlin2002}. The method is based on RecA-mediated homologous recombination occurring between homologous regions on the chromosome and plasmid. Backbone plasmid pTOF24 \citep{Merlin2002} was digested with \textit{SalI} and \textit{PstI} restriction enzymes. Inserts were amplified with primers listed in SI Table \ref{Table 1} and assembled together in Gibson assembly reaction. MG1655 was transformed with a resulting pTOF-\textit{fliC\textsuperscript{sticky} }(SI Fig. \ref{SI plasmids}A). The gene replacement was then performed following the PMGR protocol from \citep{Merlin2002}. Resulting strain EK01 was transformed with pTOF-\textit{pHluorin} plasmid (SI Fig. \ref{SI plasmids}B) and the protocol was repeated. Obtained EK07 strain differs from parental MG1655 strain in having "sticky" flagella and \textit{pHluorin} gene on the \textit{att}Tn7 site of the chromosome, which is confirmed by sequencing.

\subsection*{Supplementary Methods: pHluorin calibration}
The \textit{in vivo} calibration of the pH sensor (pHluorin) is performed as described in \citep{Wang2018}. Briefly, MM9 medium is adjusted to a set of pH values in the range between 5.5 and 9 and supplemented with 40~mM potassium benzoate and 40~mM methylamine hydrochloride \citep{Martinez2012}. The medium of a known pH is flushed into the tunnel-slide with cells attached to the surface as described before and incubated for 5~min. Total of $\sim$100 cells are imaged at 50~ms exposure time. The calibration curve obtained is given in SI Fig. \ref{SI cal curves}A and fitted with a sigmoid function  $R_{395/475}=(a_1 e^{k(pH-pH_0)}+a_2)/(e^{k(pH-pH_0)}+1)$, where $a_1, a_2, k, pH_0$ are fitting parameters. The \textit{in vitro} calibration (SI Fig. \ref{SI cal curves}B) is done using his-tagged pHluorin, purified with affinity chromatography \citep{Urh2009} diluted into buffer of known pH (and supplemented with indole or butanol when testing pHluorin sensitivity to it). The pHluorin emission intensity is measured for 395~nm and 475~nm excitation in Spark 10M multimode plate reader (Tecan Trading AG, Switzerland). pH of the buffers in presence of indole/butanol is confirmed with pH meter (FE20 FiveEasy\textsuperscript{TM}, Mettler-Toledo International Inc, Switzerland). To account for photobleaching, MM9 medium supplemented with 5\% ethanol is supplied (at 10~$\mu$l/min flow rate) to the cells attached to the surface in a flow-cell. After 5 to 10~min incubation imaging of ethanol treated cells is performed as during the photodamage experiment. The change in pHluorin intensities ratio is determined for ethanol treated cells and used to account for photobleaching by calculating the normalisation coefficients for each time point. 

\subsection*{Supplementary Methods: pHluorin image analysis}
To analyse images of pHluorin expressing cells we first identify cells uniformly attached to the surface (so called "flat cells" \citep{Pilizota2012,Buda2016}), and calculate the background intensity by dividing the whole image into four identical squares. Next we find the minimal intensity in each square and average 3x3 pixel box surrounding the minimal intensity pixel to get the background intensity value for that square. We subtract the background intensity of the corresponding square based on the location of the "flat" cell. We create a cell mask for flat cells using Otsu thresholding method \citep{Otsu1979} followed by binary erosion and binary dilation to eliminate noise. We obtain the intensities for 475~nm and 395~nm channels by calculating the mean pixel intensity within the identified cell mask, and calculate the ratio of the two to obtain the cytoplasmic pH from the \textit{in vivo} calibration curve (see above and SI Fig. \ref{SI cal curves}A).

\newpage
\subsection*{Supplementary Tables}
\begin{longtable}{ p{1.2cm}   p{3cm}   p{3.cm}  p{6cm} } 

	\hline
    & & &\\
	\textbf{Plasmid} & \textbf{Fragment name} & \textbf{Template} & \textbf{Gibson assembly primers} \\
     & & &\\
	\hline
    \hline
    & & &\\
    &\textit{fliA} shoulder & MG1655  &  5'- CCGCTTATGTCTATTGCTGGTCTCGGTACCCGACCTGCACAATGCTTCGTGACGCACCA \\ & & & 5'- AGCAGGTTCTGTCTCTGCTGCAGGGTTAATCGTTGTAACCTGATTAACTGAGACTGA\\
	& & &\\
    pTOF-\textit{fliC\textsuperscript{sticky}}&\textit{fliC\textsuperscript{sticky}} & pFD313 \citep{Berg1993} & 5'- CGTCAGTCTCAGTTAATCAGGTTACAACGATTAACCCTGCAGCAGAGACAGAACCTGCT\\
    \\ & & & 5'- CAACGACTTGCAATATAGGATAACGAATCATGGCACAAGTCATTAATACCAACAGCCTC\\ 
    & & &\\
    
    &\textit{fliD} shoulder & MG1655  & 5'- GAGGCTGTTGGTATTAATGACTTGTGCCATGATTCGTTATCCTATATTGCAAGTCGTTG\\
    \\ & & & 5'- GCTACAGGGCGCGTCCCATTCGCCACCGGTCGAAAGTTTAGCGGTAAACGACGATTG\\
    & & &\\
    
	\hline
    & & &\\
   & Tn7 Left Shoulder &MG1655  & 5’- TATGTCTATTGCTGGTCTCGGTACCCGACCTGCAATGCCGGTTATTGTTGTTGCACCGA\\ 
   \\ & & & 5’- TCGAAAGACTGGGCCTTTCGTTTTATCTGCCCGCTTACGCAGGGCATCCATTTATTACT\\
   & & &\\
    pTOF-\textit{pHluorin} &\textit{V. harveyi} promoter& pWR20 \citep{Pilizota2012} & 5’- AGTGAAAAGTTCTTCTCCTTTACTCATATGTATATCTCCTTAACTAGGTAATTATCAAGC\\
    \\ & & & 5’- ATGTTTGATTAAAAACATAACAGGAAGAAAAATGCCCCGCATTTCGACACCTTCGTCCTC\\ 
    & & &\\
    &\textit{pHluorin}& pkk223-3/pHluorin \citep{Morimoto2011}& 5’- GTAATAAATGGATGCCCTGCGTAAGCGGGCAGATAAAACGAAAGGCCCAGTCTTTCGAC\\ 
    \\ & & & 5’- TGATAATTACCTAGTTAAGGAGATATACATATGAGTAAAGGAGAAGAACTTTTCACTGGA\\ 
    & & &\\
    
      pTOF-\textit{pHluorin}&Tn7 Right  Shoulder&MG1655 & 5’- CACTTACCTGAGGACGAAGGTGTCGAAATGCGGGGCATTTTTCTTCCTGTTATGTTTTTA\\
      & &  & 5’- CAGGGCGCGTCCCATTCGCCACCGGTCGACAAACACAGAGAAAGCACTCATCGATAAGG\\ 
    & & &\\
    \hline
    \caption{List of primers used to generate pTOF24 derivatives for PMGR method.}
    \label{Table 1}
    \end{longtable}
 
\newpage
\subsection*{Supplementary Figures}

\begin{suppfigure}[h!]
\centering
 \begin{tabular}{@{}c@{}}
\includegraphics[width=0.5\linewidth]{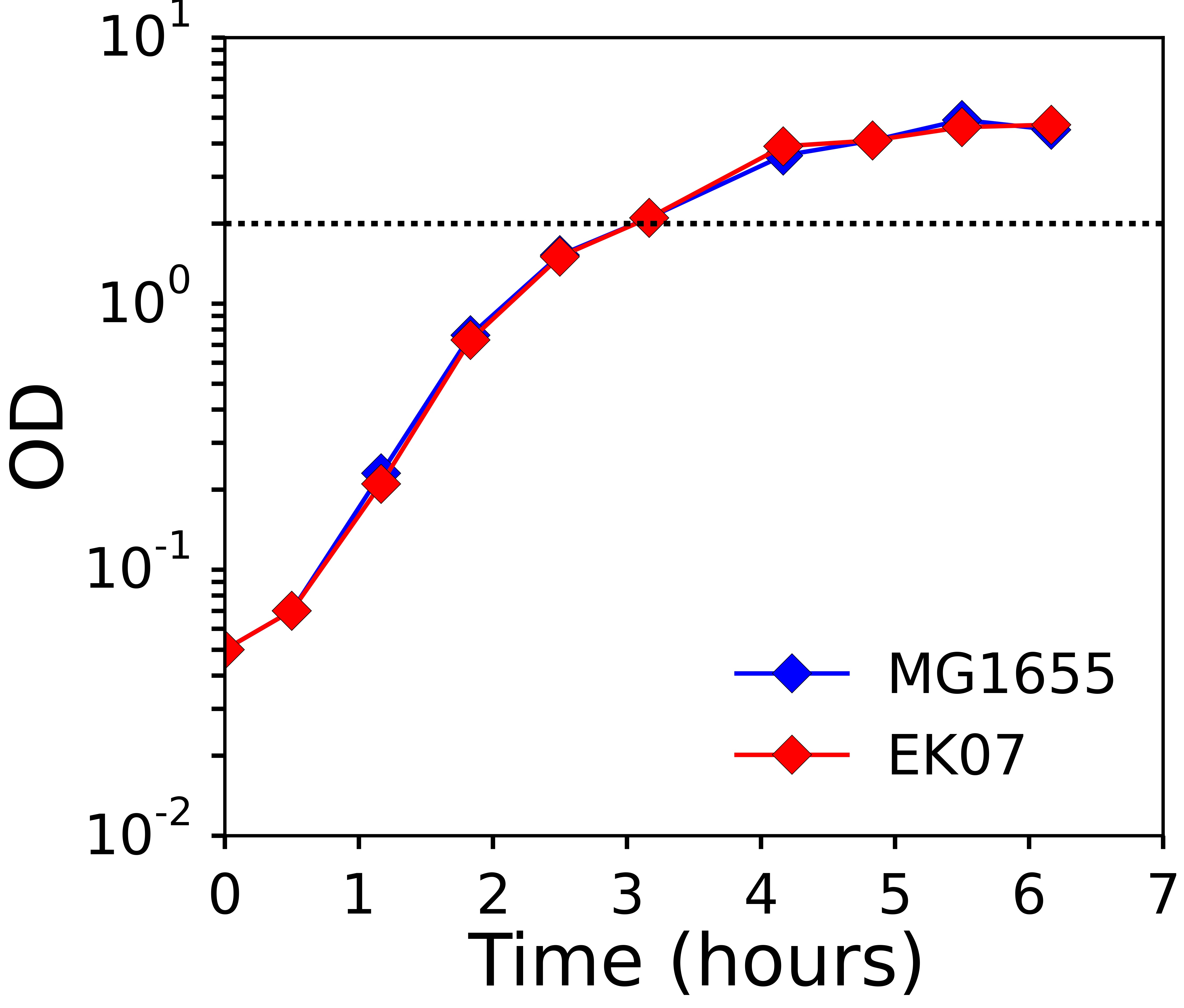}
 \end{tabular}
 \captionsetup{justification=justified}
\caption{MG1655 and EK07 growth curves in LB diluted from the overnight culture to the initial OD 0.05. Cells are grown in 125~ml flasks at 37$\degree$C while shaken at 220~rpm. Dotted line indicates OD=2, at which cells are taken for the measurements as described in \textit{Materials and Methods}.}
\label{SI growth curves}
\end{suppfigure}

\begin{suppfigure}[h!]
\centering
 \begin{tabular}{@{}c@{}}
\includegraphics[width=0.9\linewidth]{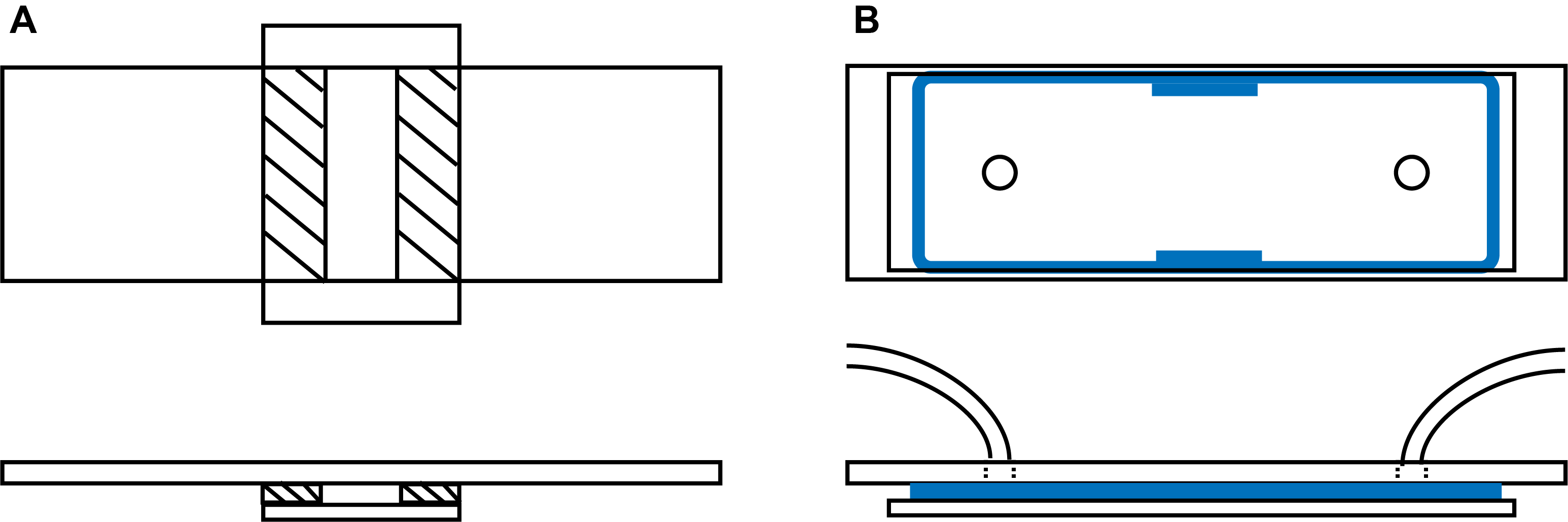}
 \end{tabular}
 \captionsetup{justification=justified}
\caption{A cartoon of the tunnel-slide and flow-cell used in the experiments. (\textbf{A}) Top view and side view of the tunnel-slide used for fast flush experiments (indole and butanol treatment): double-sided sticky tape forms a channel and is sandwiched between the microscope slide and the cover glass. Overall volume of the tunnel-slide is 10-15~$\mu$l (26x5x0.1~mm). (\textbf{B}) Top view and side view of the flow-cell used for photodamage experiments: two 1.8~mm holes are drilled in the microscope slide and connected to the tubing. The experimental chamber is formed by the gene frame and 22x60~mm cover glass. Overall volume of the flow-cell is $\sim$175-200~$\mu$l.}
\label{SI setup}
\end{suppfigure}

\begin{suppfigure}[h!]
\centering
 \begin{tabular}{@{}c@{}}
\includegraphics[width=1\linewidth]{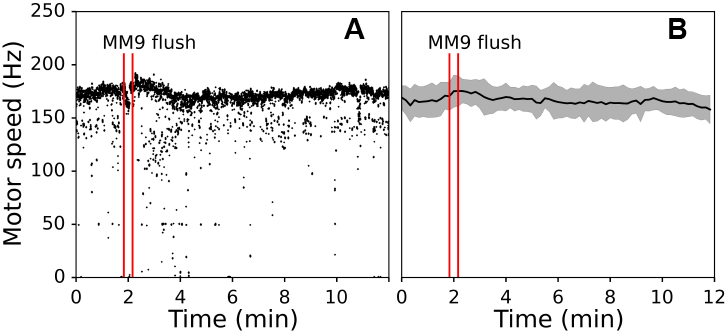}
 \end{tabular}
 \captionsetup{justification=justified}
\caption{Solely exchanging media does not affect the motor speed. (\textbf{A}) Motor speed of a cell in the tunnel-slide is recorded for 2~min, at which point fresh MM9 is flushed into the tunnel (red vertical lines indicate the duration of the flush). (\textbf{B}) Mean trace with standard deviation (shaded) of 9 cells exposed to the MM9 exchange as in (\textbf{A}).}
\label{SI flush control}
\end{suppfigure}

\begin{suppfigure}[h!]
\centering
 \begin{tabular}{@{}c@{}}
\includegraphics[width=1\linewidth]{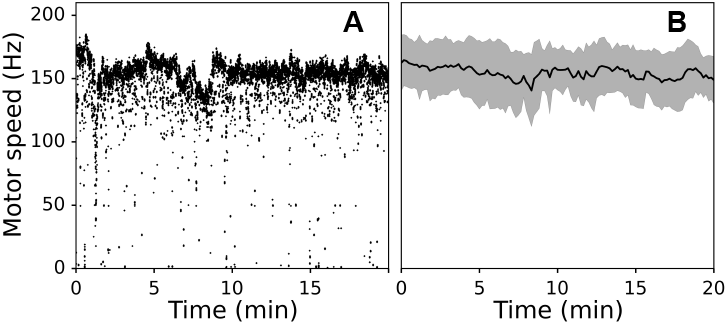}
 \end{tabular}
 \captionsetup{justification=justified}
\caption{Laser power used for back-focal-plane interferometry does not cause PMF damage. (\textbf{A}) Motor speed of a cell in a flow-cell with MM9 continuously exchanged at a 10~$\mu$l/min rate, recorded for 20~min. (\textbf{B}) Mean BFM speed trace of 17 cells monitored as in (\textbf{A}). Standard deviation is shown as shaded grey area. The control experiments were previously done for up to 45~min \citep{Rosko2017}}
\label{SI control}
\end{suppfigure}

\begin{suppfigure}[h!]
\centering
\begin{tabular}{@{}c@{}}
  \includegraphics[width=0.9\linewidth]{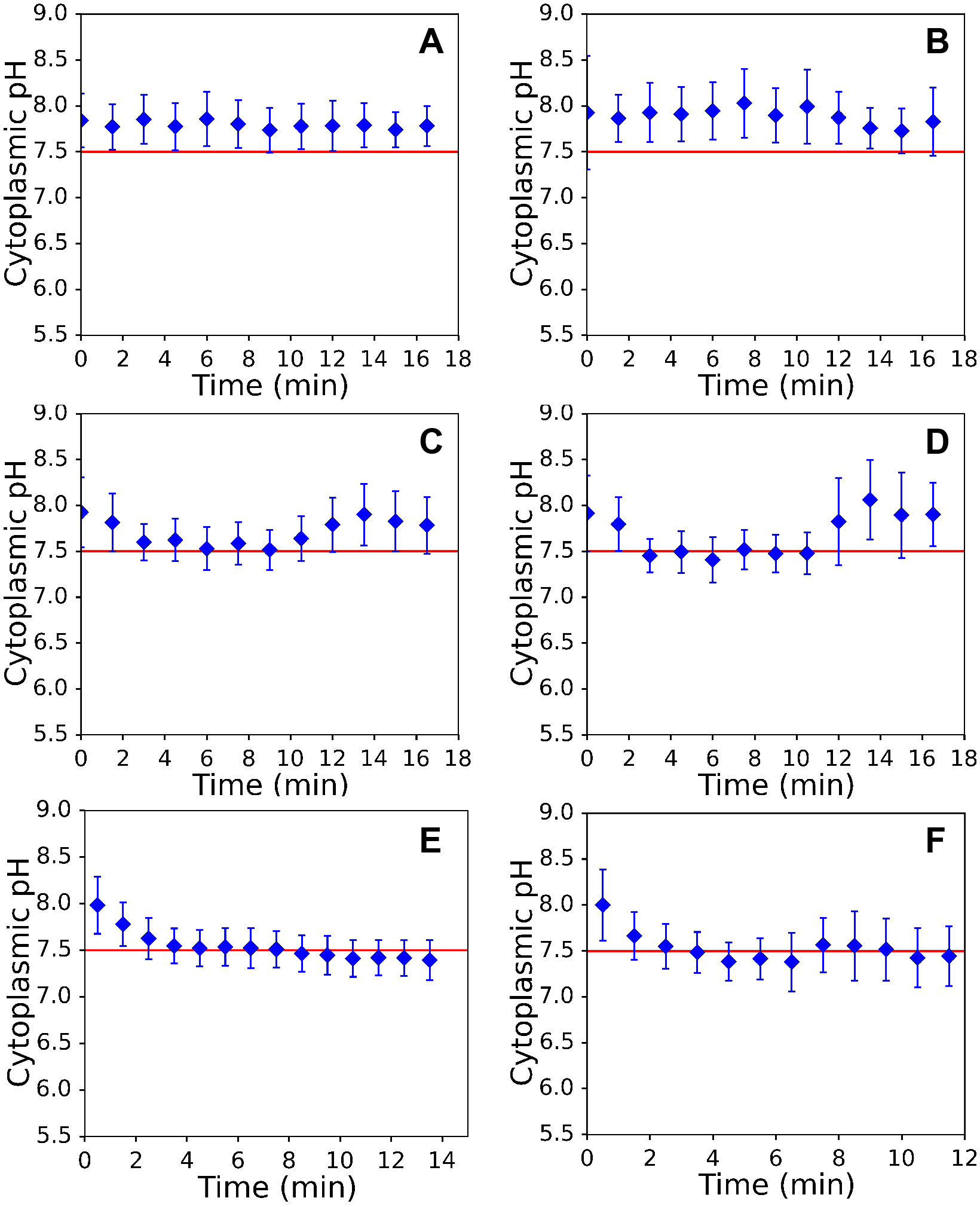}
    \end{tabular}
\captionsetup{justification=justified}
\caption{Cytoplasmic pH measurements during stress treatment. (\textbf{A}) 0.5~mM and (\textbf{B}) 2.5~mM indole, (\textbf{C}) 0.5\% and (\textbf{D}) 1\% butanol, (\textbf{E}) 153~mW/$cm^2$ and (\textbf{F}) 591~mW/$cm^2$ effective light power. Error bars represent standard deviation, and the red line value of the external pH=7.5, which we set according to the previous measurements \citep{Wang2018}. Throughout the measurements $\Delta pH \sim$0, where the cytoplasmic pH before shock in our buffers is slightly higher then previously reported (up to maximum of $\Delta pH \sim$0.5). Difference in 0.5 pH units gives $\sim$ 30~mV of PMF, which is within the standard deviation of the motor speed measurements (Fig. \ref{fig2}D or \ref{fig3}D).}
\label{SI pH}
\end{suppfigure}

\begin{suppfigure}[h!]
\centering
 \begin{tabular}{@{}c@{}}
  \includegraphics[width=0.75\linewidth]{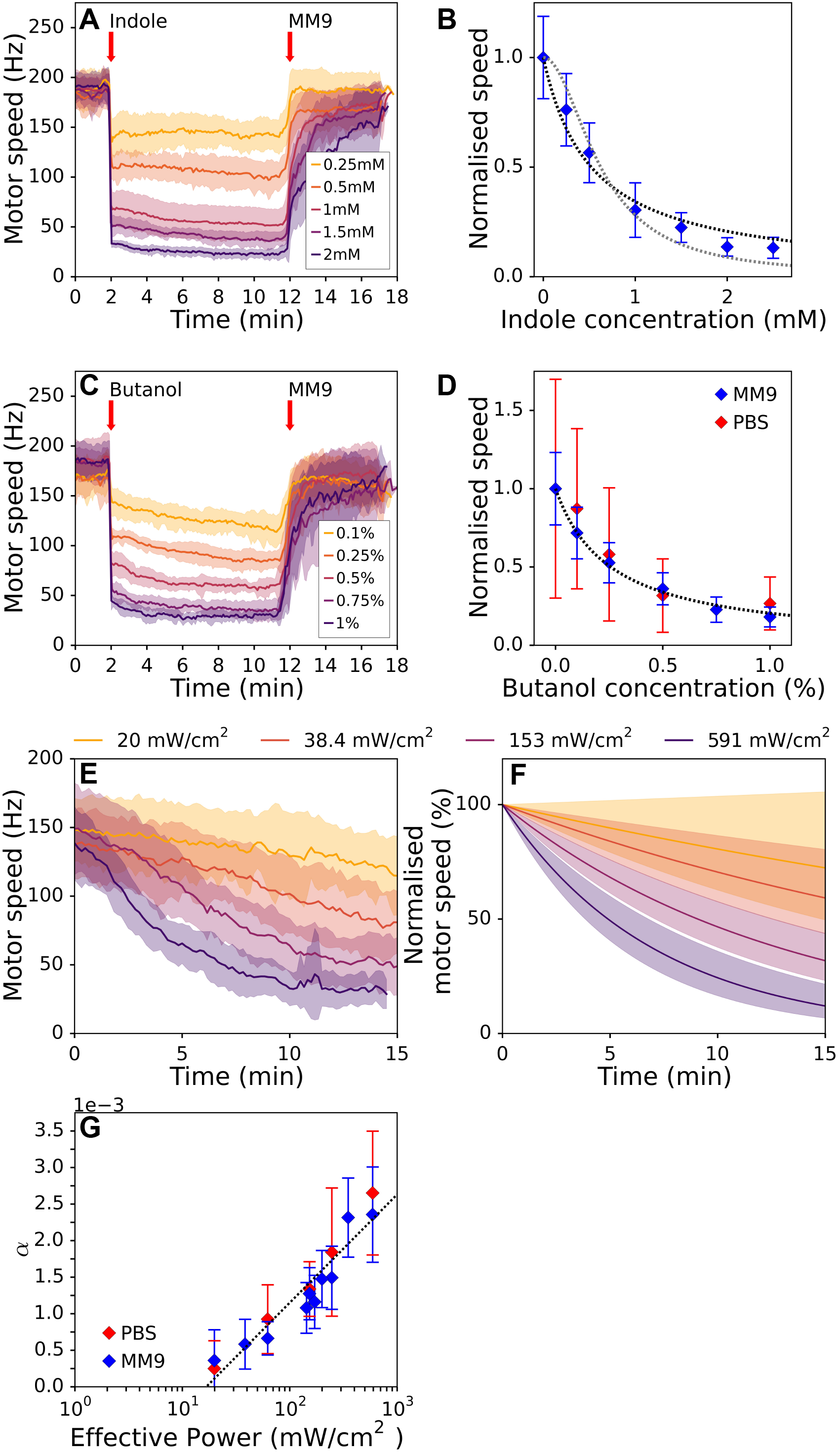}
    \end{tabular}
\captionsetup{justification=justified}
\caption{Average motor speed traces with standard deviations for all the stresses. (\textbf{A}) Version of main text Fig. \ref{fig2}B, (\textbf{B}) Fig. \ref{fig2}D, (\textbf{C}) Fig. \ref{fig3}B, (\textbf{D}) Fig. \ref{fig3}D, (\textbf{E}) Fig. \ref{fig:4}B, (\textbf{F}) Fig. \ref{fig:4}C, and (\textbf{G}) Fig. \ref{fig:4}D, but with standard deviations instead or standard errors. In (\textbf{D}) and (\textbf{G}) red diamonds represent cells in phosphate buffer saline, and blue cells in MM9. Dotted lines in (\textbf{B}), (\textbf{D}) and (\textbf{G}) are fits described in the main text.}
  \label{SI std}
\end{suppfigure}

\begin{suppfigure}[h!]
\centering
\begin{tabular}{@{}c@{}}
  \includegraphics[width=1\linewidth]{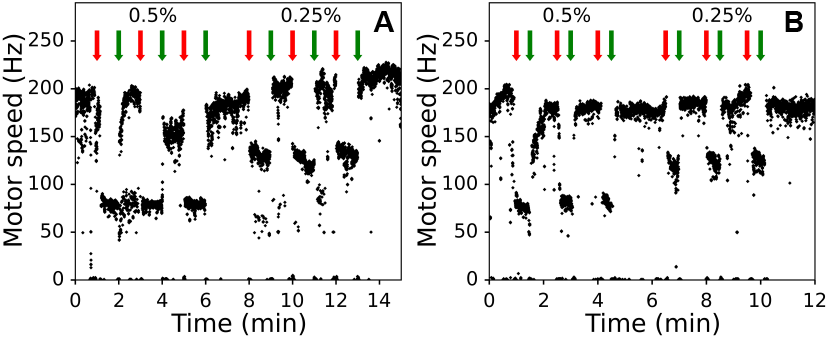}
    \end{tabular}
\captionsetup{justification=justified}
\caption{Motor speed is reversible during multiple butanol shocks. Butanol is flushed in at the time denoted with red, and out with green arrow. First three treatments are done with 0.5\%, and subsequent three with 0.25\% butanol. Intervals between butanol and MM9 flushes are of the (\textbf{A}) 1~min, or (\textbf{B}) 30~s duration.}
\label{SI multiflush}
\end{suppfigure}

\begin{suppfigure}[h!]
\centering
\begin{tabular}{@{}c@{}}
  \includegraphics[width=1\linewidth]{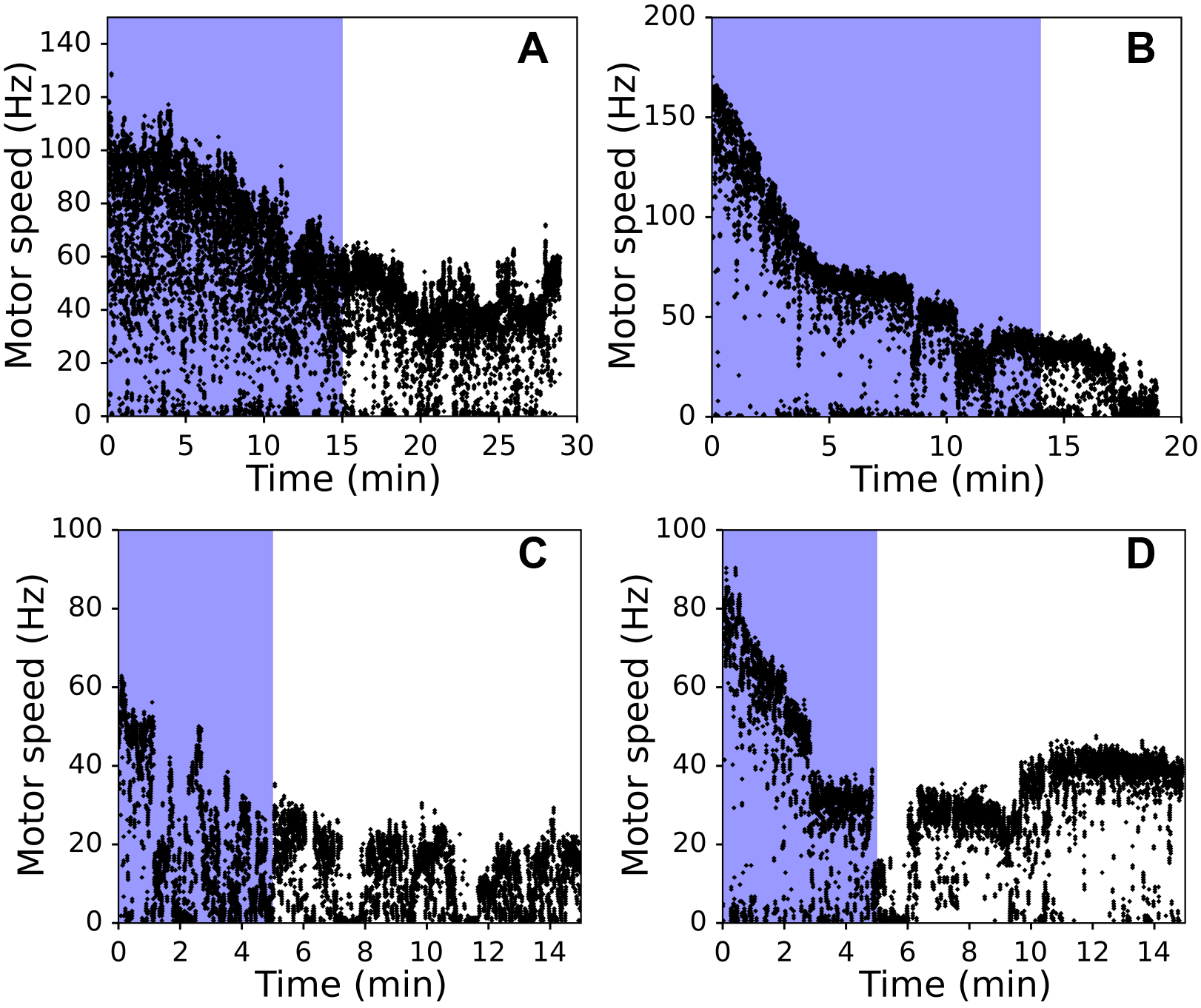}
    \end{tabular}
\captionsetup{justification=justified}
\caption{Raw single motor speed traces during and after light exposure showing that the damage is not reversible on the time scale of our experiment. Shaded blue regions indicate periods of time when the light was on. Cells are kept in MM9 in (\textbf{A}) and (\textbf{B}) and in PBS in (\textbf{C}) and (\textbf{D}). Effective light power is 38~mW/$cm^2$ for (\textbf{A}), 350~mW/$cm^2$ for (\textbf{B}) and 591~mW/$cm^2$ for (\textbf{C}) and (\textbf{D}). }
\label{SI recovery}
\end{suppfigure}

\begin{suppfigure}[h!]
\centering
 \begin{tabular}{@{}c@{}}
  \includegraphics[width=0.5\linewidth]{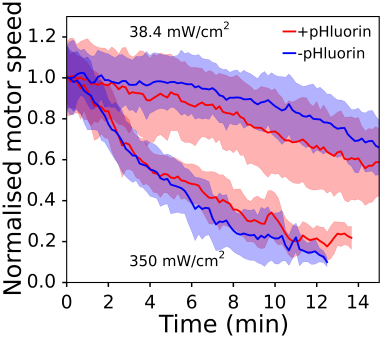}
    \end{tabular}
\captionsetup{justification=justified}
\caption{The presence of pHluorin does not alter the response to light induced damage. Averaged and normalised motor speed traces (n$\geqslant$20 cells) during exposure to light with 38.4 and 350~mW/cm\textsuperscript{2} effective power. Red traces are obtained with EK07 strain expressing pHluorin protein, and blue traces with the same strain lacking the \textit{pHluorin} gene (EK01). Shaded regions represent standard deviations.}
\label{SI fluorophore}
\end{suppfigure}

\begin{suppfigure}[h!]
\centering
\captionsetup{justification=justified}
\begin{tabular}{@{}c@{}}
  \includegraphics[width=0.8\linewidth]{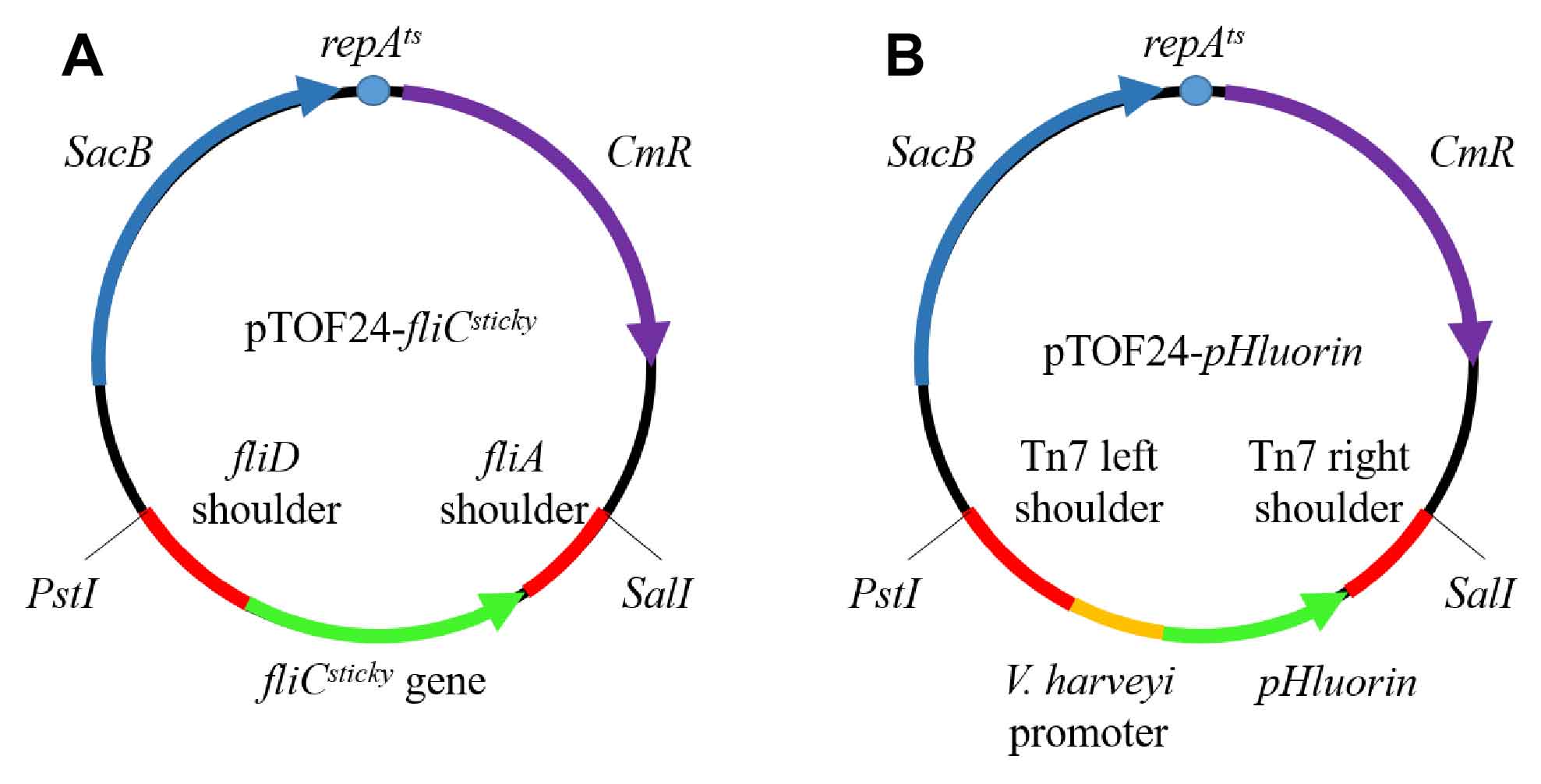}
    \end{tabular}

\caption{Maps of plasmids used to construct EK07 strain. Plasmid backbone contains chloramphenicol resistance marker (\textit{CmR}), gene encoding levansucrase (\textit{SacB}), which makes host cells sensitive to sucrose, and temperature sensitive \textit{repA\textsuperscript{ts}} replicon. (\textbf{A}) Map of pTOF-\textit{fliC\textsuperscript{sticky}}. \textit{fliC\textsuperscript{sticky}} gene flanked by $\sim$400bp sequence of homology to the surrounding chromosome (here named \textit{fliA} and \textit{fliD} shoulders) is cloned into pTOF24 between \textit{PstI} and \textit{SalI} cloning sites. Resulting plasmid has a chloramphenicol resistance. (\textbf{B}) Map of pTOF-\textit{pHluorin}. \textit{pHluorin} gene under strong constitutive \textit{V. harveyi} cytochrome C oxidase promoter \citep{Pilizota2012} flanked by $\sim$350bp sequence of homology to the \textit{att}Tn7 site on the chromosome is cloned into pTOF24 between \textit{PstI} and \textit{SalI} cloning sites.}
\label{SI plasmids}
\end{suppfigure}

\begin{suppfigure}[h!]
\centering
\includegraphics[width=\linewidth]{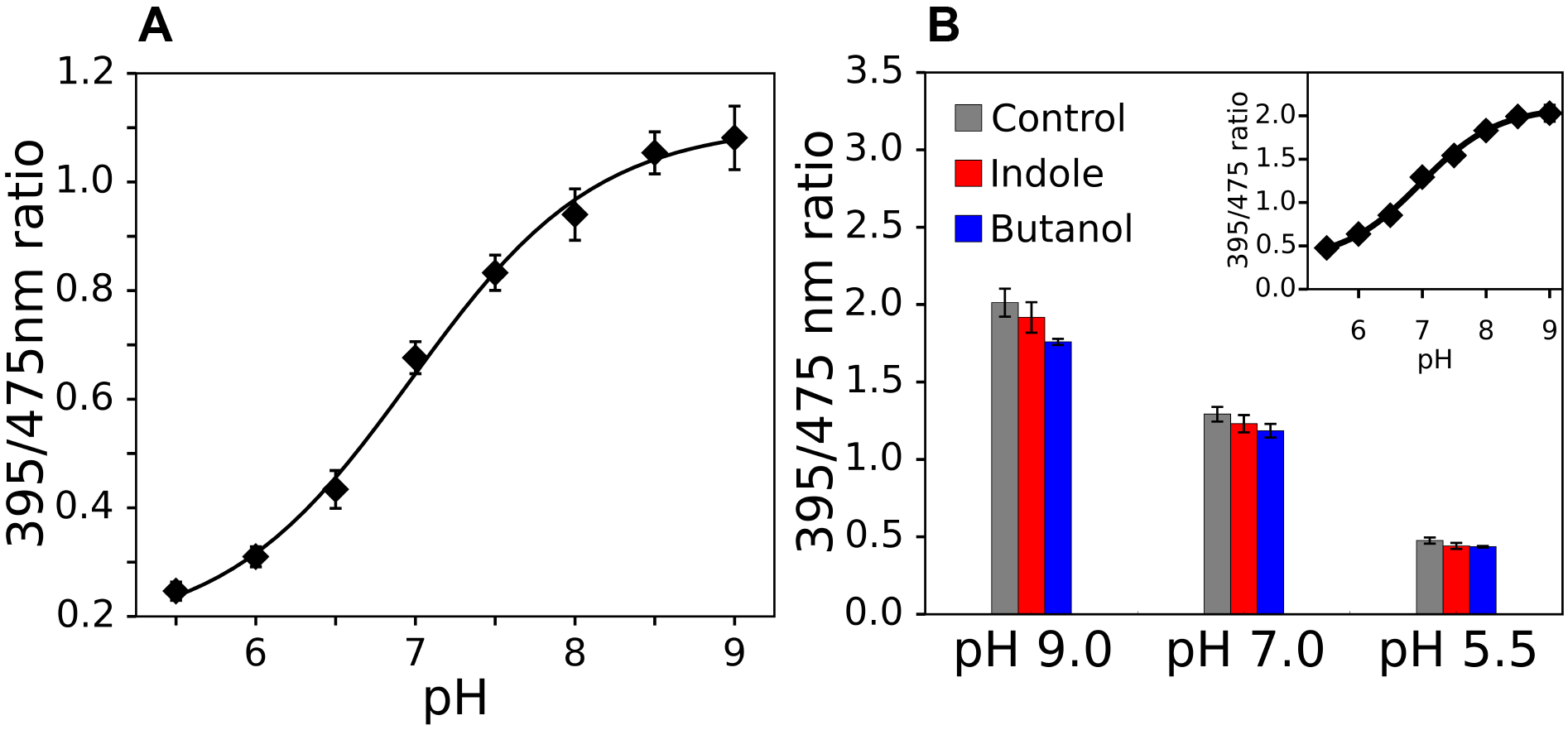}
\captionsetup{justification=justified}
\caption{pHluorin is calibrated \textit{in vivo} and \textit{in vitro}. (\textbf{A}) \textit{In vivo} calibration curve is obtained as described in \textit{SI Materials and methods}. 395/475~nm emission intensity ratio plotted against pH with standard deviation. The experimental points are fitted with a sigmoid function $R_{395/475}=(a_1 e^{k(pH-pH_0)}+a_2)/(e^{k(pH-pH_0)}+1)$, where $a_1=1.1$, $a_2=0.168$, $k=1.7$ and $pH_0=6.976$ are obtained from the fit. (\textbf{B}) 395/475~nm emission intensity ratio of purified pHluorin in presence of butanol and indole at different pH; error bars are standard deviation. pHluorin readings are influenced by the presence of indole/butanol for pH 9, but not pH 7 (or 5.5) at which we perform our experiments. \textbf{Inset}: Calibration curve of the purified pHluorin. Parameters obtained from fitting the experimental data to the sigmoid function are: $a_1=2.1$, $a_2=0.33$, $k=1.7$ and $pH_0=6.97$.}
\label{SI cal curves}
\end{suppfigure}

\end{document}